**Perceptual and technical barriers in sharing and formatting metadata accompanying omics studies**

Authors:

Yu-Ning Huang
Department of Clinical Pharmacy, Alfred E. Mann School of Pharmacy and Pharmaceutical Sciences, University of Southern California, Los Angeles, California, 90089, USA
yuninghu@usc.edu
ORCID: 0000-0003-1697-4267

Michael I. Love
Department of Genetics, University of North Carolina at Chapel Hill, Chapel Hill, NC, 27514, USA
Department of Biostatistics, University of North Carolina at Chapel Hill, Chapel Hill, NC, 27514, USA
michaelisaiahlove@gmail.com
ORCID: 0000-0001-8401-0545

Cynthia Flaire Ronkowski
Department of Clinical Pharmacy, Alfred E. Mann School of Pharmacy and Pharmaceutical Sciences, University of Southern California, Los Angeles, California, 90089, USA
cynthiaflaire@gmail.com

Dhrithi Deshpande
Department of Clinical Pharmacy, Alfred E. Mann School of Pharmacy and Pharmaceutical Sciences, University of Southern California, Los Angeles, California, 90089, USA
dhrithideshpande@gmail.com
ORCID: 0000-0003-3794-7364

Lynn M. Schriml
Department of Epidemiology and Public Health, Institute for Genome Sciences, University of Maryland School of Medicine, Baltimore, Maryland 21201, USA
lschriml@som.umaryland.edu
ORCID: 0000-0001-8910-9851
Twitter: @lschriml

Annie Wong-Beringer
Department of Clinical Pharmacy, Alfred E. Mann School of Pharmacy and Pharmaceutical Sciences, University of Southern California, Los Angeles, California, 90089, USA

anniew@usc.edu
ORCID: 0000-0003-3302-1409

Barend Mons
Leiden University Medical Center, Leiden Academic Centre for Drug Research, GO FAIR foundation, Leiden, The Netherlands
barendmons@gmail.com
ORCID: 0000-0003-3934-0072

Russell Corbett-Detig
Department of Biomolecular Engineering, University of California, Santa Cruz. Santa Cruz, CA 95064.
Genomics Institute, University of California, Santa Cruz. Santa Cruz, CA 95064
rucorbet@ucsc.edu
ORCID: 0000-0001-6535-2478

Christopher I Hunter
GigaDB Director, GigaScience Press, Hong Kong
chris@gigasciencejournal.com
ORCID: 0000-0002-1335-0881

Jason H. Moore
Department of Computational Biomedicine, Cedars-Sinai Medical Center, Los Angeles, CA 90069
Jason.Moore@csmc.edu
ORCID: 0000-0002-5015-1099

Lana X. Garmire
Department of Computational Medicine and Bioinformatics, Medical School, University of Michigan, Ann Arbor, MI 48105
lgarmire@med.umich.edu
ORCID: 0000-0003-1672-6917
Twitter: @GarmireGroup

T.B.K. Reddy
DOE Joint Genome Institute, Lawrence Berkeley National Laboratory, Berkeley, CA 94720, USA
TBReddy@lbl.gov
ORCID: 0000-0002-0871-5567

Winston A. Hide
Department of Pathology, Beth Israel Deaconess Medical Center, Harvard Medical School, 330 Brookline Avenue
Boston, MA 02215-5491
ORCID: 0000-0002-8621-3271
whide@bidmc.harvard.edu

Atul J. Butte
atul.butte@ucsf.edu
Bakar Computational Health Sciences Institute, University of California, San Francisco, San Francisco, CA 94143 USA
Center for Data-Driven Insights and Innovation, University of California, Office of the President, Oakland, CA 94607 USA
ORCID: 0000-0002-7433-2740
Twitter: @atulbutte

Mark D. Robinson
SIB Swiss Institute of Bioinformatics and Department of Molecular Life Sciences, University of Zurich, 8057 Zurich, Switzerland
ORCID: 0000-0002-3048-5518
mark.robinson@mls.uzh.ch

Serghei Mangul
Department of Clinical Pharmacy, Alfred E. Mann School of Pharmacy and Pharmaceutical Sciences, University of Southern California, Los Angeles, CA 90033, USA
Department of Quantitative and Computational Biology, USC Dornsife College of Letters, Arts and Sciences, University of Southern California, Los Angeles, CA 90033, USA
mangul@usc.edu
ORCID: 0000-0003-4770-3443

**Abstract**

Metadata, often termed "data about data," is crucial for organizing, understanding, and managing vast omics datasets. It aids in efficient data discovery, integration, and interpretation, enabling users to access, comprehend, and utilize data effectively. Its significance spans the domains of scientific research, facilitating data reproducibility, reusability, and secondary analysis. However, numerous perceptual and technical barriers hinder the sharing of metadata among researchers. These barriers compromise the reliability of research results and hinder integrative meta-analyses of omics studies . This study highlights the key barriers to metadata sharing, including the lack of uniform standards, privacy and legal concerns, limitations in study design, limited incentives, inadequate infrastructure, and the dearth of well-trained personnel for metadata management and reuse. Proposed solutions include emphasizing the promotion of standardization, educational efforts, the role of journals and funding agencies, incentives and rewards, and the improvement of infrastructure. More accurate, reliable, and impactful research outcomes are achievable if the scientific community addresses these barriers, facilitating more accurate, reliable, and impactful research outcomes.

**The power of metadata in multi-omics data analysis**

Over the last decade, advancements in next-generation sequencing technologies have democratized access to a vast array of public omics data across disparate diseases and phenotypes[1]. Typically, public multi-omics data are widely available and discoverable in public repositories[2–4] such as the ArrayExpress[2], Gene Expression Omnibus[4] (GEO) and the Sequence Read Archive[3] (SRA). These public repositories serve as important platforms for storing multi-omics data and accompanying metadata, generated from a diverse array of studies. Metadata refers to the descriptive and contextual information about the generation, provenance, and context of raw data, including experimental design, instrumentation parameters, and data processing steps. Importantly, ensuring that metadata accompanying raw omics data adheres to the FAIR (Findable, Accessible, Interoperable, and Reusable) principles is crucial in order to establish a comprehensive framework for data management[5–7]. By incorporating the principles of FAIR, the data not only becomes more discoverable and available but also becomes capable of undergoing seamless cross-examination through distributed analytics and learning across research domains. When attempting cross-domain analyses of data, interoperability becomes critical to alleviate disparate vocabularies and conceptual models.

In particular, metadata plays a crucial role in data management and analysis. It provides the crucial context that helps researchers understand, manage, manipulate, and analyze omics data[8,9]. Its value lies in how people (and increasingly machines) utilize it to enhance their "understanding" of data sources. Metadata aids in locating the specific types of data required, making searches more efficient and targeted. It contributes to result interpretation and explainability, allowing users to comprehend and communicate the underlying processes and factors influencing outcomes. In the realm of databases, metadata enables efficient organization and retrieval of data, facilitating seamless access and analysis. For the purpose of discoverability, and for many other reasons, it is considered good practice in the FAIR context to separate the FAIR metadata from the actual data they describe. Even if the data itself may be FAIR but restricted, for instance because it is person-sensitive, the metadata may be open. In addition, data that are not necessarily FAIR themselves can be made machine actionable, meaning that the data is organized and formatted in a manner that enables automated processing, typically through programming or algorithms, without the need for human interpretation[10,11]. This machine actionable approach enables efficient automated analysis, retrieval, and utilization of data, even if the data is not inherently FAIR on its own. This is why the use of FAIR standards to structure metadata and data is crucial in the era of data-intensive and machine-assisted science[12]. Comprehensive metadata documentation, paired with raw omics data, plays a pivotal role in promoting reproducibility. It enables accurate replication of research, experiments, or analyses, facilitating the assessment of preprocessing and modeling choices, thereby enhancing scientific rigor[13,14]. Overall, by harnessing the power of metadata, researchers can unlock numerous benefits of datasets, otherwise unavailable, that enhance data understanding, search capabilities, interpretation, database management, and reproducibility.

**The role of metadata in secondary analysis**

Secondary analysis, the re-analysis of existing data and metadata, is a powerful research approach that can lead to novel biomedical discoveries across life sciences[15,16]. Accurate and well-structured metadata are vital for effective secondary analysis[17]. For example, leveraging metadata like age, sex, and disease conditions enables precise integration and comparison of results across diverse studies[9,16]. This crucial combination ensures accurate secondary analyses, forming a robust foundation for profound insights[1,16]. Searchable, findable, and well-curated metadata can spark new projects and discoveries. An example of how curated metadata resources impact discovery is Genomes OnLine Database (GOLD)[18], which has helped in research leading to new publications. The curated ecosystem metadata from GOLD has helped the authors to determine the distribution of the PHA synthase (PhaC) genotype in different environments and help them tabulate different PhaCs in different environments. Serratus, a petabase-scale sequence alignment resource developed by Edgar et. al.[19], it integrated curated virus host metadata that helped to characterize novel viruses and their environmental reservoirs. Further, organizing metadata with controlled vocabularies, ontologies and standardized classifications enabled further new discoveries. For example, Vuong et. al.[20] performed a large-scale mining of microbial genomes to develop bioprospecting strategies for bioplastics, a task that was made possible by the use of standardized metadata and ontologies. This study exemplifies the power of well-structured metadata in enabling new scientific insights and accelerating the pace of discovery.

**The need for improved metadata sharing practices**

Scientific journals and research organizations enforce sharing of raw omics data via guidelines and policies[5,6,21–24], but guidance on metadata sharing is limited[25]. A survey of 506 neuroscientists found only 33% embraced standardized data sharing guidelines[26]. Omics data, with their high dimensionality and diverse types, pose interpretation challenges for secondary analyses that combine multiple sources, given each dataset's unique challenges and metadata needs[27,28]. For instance, variations in ontologies and annotations used to characterize proteins and genes can hinder the seamless use of omics data across studies or in secondary analysis. Publishing a detailed study description, methodology, results, and interpretation is crucial. Making all research products, including data (where possible) and corresponding metadata, FAIR (Findable, Accessible, Interoperable, Reusable), well-documented, and organized is essential for reproducible, efficient, and accurate secondary analyses. While some data cannot be public due to confidentiality[29,30], sharing metadata—providing information on data existence, characteristics, and potential access restrictions—is encouraged and should be linked to the actual data. Data transparency and availability, coupled with accessible metadata, enhance reproducibility and the robustness of scientific research in the era of data-intensive projects[30].

**Overcoming barriers to metadata sharing**

Perceptual and technical obstacles can prevent research scientists from sharing metadata[16,31,32] and lead to challenges in integrative meta-analysis of omics data across multiple cohorts to compromise the reliability of the data[16]. For instance, one barrier includes insufficiently detailed metadata for critical aspects of the experimental units, such as population descriptors (race, ethnicity, ancestry), age, disease condition, and, importantly, sex[16,33]. Additionally, there is a need for the appropriate use of population descriptors [34,35]. In the absence of important metadata researchers would not be able to accurately leverage published raw sequencing data for secondary analysis, e.g., if ancestry information is missing[36]. By identifying and addressing barriers to metadata sharing practices, future researchers can ensure the availability, completeness, and accuracy of metadata. Below, we outline the existing barriers impeding metadata sharing practices among researchers and propose potential solutions to overcome such obstacles.

**The insufficient adoption of uniform standards and guidelines makes it challenging for researchers to report complete, standardized, and high-quality metadata.**

The insufficient adoption of standards, such as FAIR compliance[5] is one of the key barriers to metadata sharing. This results in non-uniform metadata practices, hindering cross-examination, limiting comprehensive database development, and complicating secondary analysis processes[26,37]. The absence of adoption of FAIR standards in public omics data across projects has resulted in diverse metadata reporting practices. For instance, sharing population information varies - some report ancestry, others ethnicity or race - introducing discrepancies, unresolved complexity and differing definitions of descriptors[36,38]. These subtle differences in definitions result in distinct clinical implications[36,38,39]. Additionally, while these standards may meet US Federal requirements, their misalignment with international standards results in the absence of globally unique identifiers, leading to significant data and metadata variations. The diverse ontologies in metadata complicate integrating large amounts of data across study cohorts, making the process time-consuming and error-prone[6,40,41]. Without FAIR metadata reporting practices, matching and aligning metadata attributes, such as experimental conditions, sample characteristics (e.g., collection date, condition of specimen), and data preprocessing methods, become complex and error-prone. Organizations, such as the Global Alliance for Genomics and Health (GA4GH)[42] and the Genomic Standards Consortium (GSC)[43], have published standards for genomic data sharing, and the Public Health Alliance for Genomic Epidemiology (PHA4GE) has published standards for genomic epidemiology[44,45]. Other groups, such as the Observational Health Data Sciences and Informatics (OHDSI)[46] and the Clinical Data Interchange Standards Consortium (CDISC)[47] also publish the data models for observational health data and clinical data. Numerous data standards and models underscore the significance of sharing data and metadata in a consistent way. However, the lack of universally accepted consensus or, at least, mandated minimal information standards for data and metadata sharing across different scientific domains leaves researchers uncertain about appropriate guidelines to follow and what information to share.

The absence of standardized metadata reporting guidelines introduces uncertainty and results in inconsistent and incomplete information across studies[41], posing challenges for integrating and analyzing samples from diverse study cohorts[48]. For example, our previous study on sepsis investigated metadata availability in raw data and identified inconsistencies in reporting tissue type information. Studies used various non-standardized formats, presenting tissue types as either "source" information or "tissue" information [16]; at the point of secondary analysis, such inconsistencies need to be resolved. Ultimately, researchers must not only report tissue information but also explicitly specify the types involved (e.g., liver biopsy or kidney biopsy) for comprehensive adherence to metadata standard guidelines. While standards often exist, the challenge lies in ensuring researcher adoption and proper implementation to advance research quality and reproducibility. In conclusion, the lack of adoption and implementation of standardized guidelines hinders the integration and interpretation of omics data across various research fields.

**Privacy, legal and ethical concerns for the biomedical communities limit metadata sharing in the public domain**

Another challenge in metadata sharing pertains to the privacy, legal, and ethical concerns of individuals who have contributed the biospecimens[36,42,49]. Metadata and/or data can contain sensitive information that, if disclosed, could potentially compromise the study participants' privacy[36,50]. As a result, data and metadata containing personally identifiable information pose a major barrier to data sharing due to privacy concerns[50]. Such data cannot and should not be shared without prior de-identification. Additionally, metadata sharing may involve legal barriers with respect to privacy protection. Stringent metadata and data sharing regulations may further hinder metadata availability[51]. Local privacy laws and regulations must be carefully considered and followed to ensure compliance with established data privacy protection guidelines and frameworks[52]. For example, the Health Insurance Portability and Accountability Act (HIPAA) is a Federal law enacted in the United States in 1996 with the primary goal of protecting the privacy and security of individuals' health information[53]. Given HIPAA's strong emphasis on protecting individuals' health data privacy, researchers with access to identity-containing metadata may face stricter authorization, data de-identification, and security measures.. These requirements can add complexity and administrative burdens, potentially deterring researchers funded by the US government agencies from sharing the US-population specific metadata. In the nearly 30 years since HIPAA was passed, many US states have enacted newer privacy laws, such as the California Consumer Privacy Act (CCPA).

Similarly, within the European Union (EU), researchers engaged in the handling and sharing of data belonging to EU citizens encounter a significant legal framework known as the General Data Protection Regulation (GDPR)[54,55]. Enacted to safeguard the privacy and rights of individuals, this comprehensive legislation imposes stringent guidelines for the collection, processing, storage, transfer, analysis, and dissemination of personal data of EU citizens. While this framework aims to enhance data protection and empower individuals with control over their personal information, it can also introduce significant legal barriers for metadata sharing. Researchers must navigate these legal intricacies to ensure that their activities align with GDPR requirements, potentially leading to limitations in metadata sharing. The GDPR's emphasis on consent, data minimization, and accountability, though vital for safeguarding EU citizens' data rights, adds an extra layer of responsibility to the research process.

Additionally, there are concerns about the possibility of data leaks or breaches when sharing metadata, which might also prevent metadata sharing practices[26,56]. As of August 2023, the Cam4 data breach in March 2020 remains the largest reported data leakage, exposing over 10 billion data records. The second-largest data breach in history, the Yahoo data breach, occurred in 2013[57]. These security concerns not only compromise the integrity of the data but also violate privacy regulations, casting doubt over the utility and safety of disseminating metadata openly. This hesitance can limit scientific collaboration, hinder the advancement of research, and slow the pace of discovery, as researchers may be reluctant to share valuable metadata without assured protections against unauthorized access or misuse. Lastly, ethical and cultural considerations also come into play when sharing metadata[36]. Some researchers may hesitate to share metadata from their studies due to cultural practice[58]. These concerns stem from various factors, including intellectual property concerns, competition, commercial reasons, or personal preferences regarding the level of transparency in sharing detailed metadata accompanying raw omics data.

**Limitations in study design prevent researchers from sharing phenotypes not approved by IRB**

The availability of metadata can be significantly constrained by the study design[16]. Several barriers hinder effective metadata collection. These begin with the lack of planning for metadata collection during the experiment design phase, such as omitting metadata collection protocols in original Institutional Review Board (IRB) applications or devising a study wide metadata collection plan prior to a multi-site soil collection event. Without adequate forethought and consideration for metadata collection, researchers may overlook crucial aspects or label the same data element in different ways resulting in incomplete or absent metadata. Additionally, an important aspect to consider revolves around the patients' perspective within the realm of IRB limitations. When conducting the initial study and securing informed consent from patients for HIPAA data usage, patients may choose not to grant consent for the perpetual utilization of their safeguarded health data. This decision would restrict usage of their data for secondary analysis in unanticipated hypothesis testing. Such limitations can have a consequential impact on the extent and feasibility of metadata sharing practices in clinical settings. Furthermore, poor data collection methods, such as non-standardized and inconsistent metadata collection, can compromise the reliability and quality of the metadata[59], leading to discrepancies in formats, units of measurement, ontology, or even the inclusion/exclusion of essential information[1]. As a result, these discrepancies can introduce bias, hinder data integration, and limit the potential insights that can be derived from the data.

**Limited incentives for researchers to share metadata**

A significant barrier to effective metadata sharing practices is the absence of motivation and incentives for researchers to allocate time and resources towards the accurate collection and sharing of metadata[25,42,60]. The paucity of incentives for researchers in sharing metadata poses challenges to the discovery and reproducibility of research results based on existing raw data. Due to the prevailing emphasis on publishing articles in high impact factor journals and the sense of "owning the data", researchers often prioritize activities directly related to manuscript preparation and publication, overlooking the importance of data and metadata sharing[60,61]. This is coupled to a pervasive lack of understanding of the value of metadata, the increased potential for citation of the article and its data, and a lack of incentives for re-use of the data. Additionally, for all academic, research, and private laboratories, questions arise about how to distribute the financial responsibility for additional costs related to training and setting up the infrastructures for data collection. As a result, research data may remain under shared and underutilized, impeding the potential for new discoveries and hindering the ability of other researchers to replicate and build upon existing findings.

**Inadequate infrastructures for sharing and storing metadata negatively affects its availability**

Insufficient infrastructure for sharing and storing metadata, along with the absence of systematic data management practices, presents significant obstacles for researchers seeking to repurpose raw data effectively[26,51,62,63]. This barrier often arises from the disconnect in the storage of metadata and the primary raw data, leading to difficulties in accessing and seamlessly integrating the information[51]. For instance, metadata may be stored in different locations such as public repositories or within the original publication[16]. Difficulties may arise from extracting metadata from publication text using Natural-Language Processing (NLP) methods[64,65] or from extracting metadata directly from public repositories using other code-based techniques[16]. The above approaches for extracting metadata poses significant technical barriers for researchers involved in secondary analysis. As a result, without mandatory metadata deposition in public archives, data sharing will not improve, regardless of the numerous data sharing policies in place. Additionally, there are notable variations in both the quality and quantity of data storage repositories among different countries[66]. These discrepancies can worsen the lack of metadata and quality issues in diverse contributing countries. The lack of sufficient metadata management systems hampers effective organization and use, hindering raw data's reproducibility and repurposing[26,62].

**Lack of well-trained personnel for systemic management for metadata negatively impacts the availability of metadata**

The inadequate training of personnel in metadata sharing can result in a range of challenges in metadata management, including the presence of inaccurate or incomplete reported metadata, an elevated risk of data breaches and data loss, and inefficient utilization of the available data resources[67]. Several barriers contribute to these issues. Firstly, metadata is often highly technical and specialized, demanding expertise in the specific field to ensure accurate interpretation and annotation[68,69]. Additionally, not all researchers possess the necessary computational training to effectively share and publish metadata alongside raw data in structured FAIR-compliant formats. Next, the lack of personnel trained in effective metadata annotation and description can lead to delays in metadata documentation and incomplete metadata records, ultimately hindering the utility and comprehensiveness of metadata for downstream research[63]. Lastly, without skilled individuals proficient in metadata management practices, there is a higher likelihood of inconsistent or incomplete metadata records, leading to difficulties in locating and utilizing relevant data. The lack of well-trained personnel thus poses a significant obstacle to ensuring the availability and usability of metadata within a system[62].

**Promoting standardization: The need for universally accepted metadata reporting guidelines**

The development and adoption of standardized metadata reporting guidelines holds immense promise for enhancing metadata availability, particularly within eukaryotic sequencing projects. Currently, reporting practices for human-associated metadata, outbreak or infectious disease-related data, and environmental microbiome data vary significantly across different communities. While standards for metadata reporting in microbial sequencing studies have been established for some time[15,70], there remains a pressing need for a comprehensive set of reporting guidelines specifically tailored to eukaryotic sequencing projects. Firstly, it is imperative that dedicated efforts are undertaken to facilitate the development and adoption of standardized metadata reporting guidelines[6,41,71]. While numerous publications and guidelines exist for metadata sharing practices, the absence of a consensus on which guidelines to follow results in a wide range of reported metadata approaches[6,22,23]. It is important to recognize that distinct types of metadata, such as those obtained from human, microbial, environmental, and others, necessitate specific metadata guidelines tailored to their respective domains. By actively investing resources, expertise, and collaboration, the scientific community can ultimately establish robust, published[5,72] guidelines that encompass the diverse requirements across domains. Well-defined guidelines are essential for ensuring that collected metadata is machine-readable, actionable, and complies with the FAIR principles. Clear documentation and guidelines outlining metadata management processes and standards should be established for easy reference. We advocate for comprehensive metadata submission, encompassing detailed study descriptions and sample information. Enhance metadata capabilities through the addition of custom fields or collaboration with standards developers to improve existing frameworks. These strategies optimize data organization and accessibility, promoting effective data management and sharing.

While establishing standards is a crucial initial step, the current bottleneck hindering the progress of the field lies in the rigorous application of these standards. This challenge serves as the primary obstacle to the widespread sharing of metadata. Overcoming this hurdle necessitates a concentrated effort on promoting the comprehensive implementation of metadata sharing guidelines with available training. A noteworthy initiative addressing this requirement is the National Microbiome Data Collaborative (NMDC)[70]. NMDC is actively dedicated to enhancing the adoption of standardized metadata practices within the microbiome research community. However, to create a substantial impact, these initiatives need to be expanded on a larger scale, reaching across diverse domains and engaging researchers on a broader scale. The metadata sharing standard should also address legal and ethical considerations for specific data types, particularly human data, across diverse jurisdictions. For example, the Nagoya Protocol, a harmonized international agreement, promotes data sharing by providing a clear framework for access to and benefit-sharing from genetic resources and traditional knowledge. It encourages transparency and equitable collaboration, building trust and facilitating data exchange[73]. Other research practices can also guide the proper metadata and data sharing practices. For instance, proper data handling practices include obtaining informed consent from study participants and using de-identification techniques to maintain the trust and ethical integrity of raw data analyses[74]. In addition, clear guidelines for metadata collection enable researchers to account for these requirements before submitting their IRB applications. Additionally, establishing which subsets of IRB-approved metadata can be shared openly facilitates the open sharing of at least non-identifiable data. The implementation of a comprehensive protocol for metadata collection, along with the maintenance of good laboratory/clinical practices (GXPs), can effectively ensure the high quality and reliability of metadata collected during experimental settings. In conclusion, implementation of metadata sharing guidelines is essential to promote effective data reuse and facilitate cross-study analysis and secondary analysis.

Another potential solution involves establishing standards for providing the minimum sample-related information. Although achieving universal consensus in scientific domains can be challenging, the Minimum Information for Biological and Biomedical Investigations (MIBBI) guidelines, developed by the FAIRsharing group, provide a standardized approach for reporting minimal information from data generated using relevant methods across various bioscience fields[7,75]. Adherence to MIBBI guidelines not only ensures transparency in reporting experiments, enhances data accessibility, and facilitates effective quality assessment but also elevates the overall value of a body of work. It further enables the creation of structured databases, public repositories, and the development of data analysis tools, instilling confidence in researchers to share research-related data[40].

**Educational efforts: Educational programs and workshops are essential to improve the quality and availability of metadata accompanying scientific research**

Educational programs and training workshops can educate researchers with the importance of metadata sharing and the technical instructions on adopting metadata sharing guidelines[6,41,71,76], equipping researchers with the necessary skills and knowledge to effectively handle metadata. These educational efforts should focus on the value and impact of proper metadata, enhancing understanding of metadata standards[76], and data management techniques[63], and ensure the quality and compatibility of metadata across different datasets[77]. Training researchers to prioritize metadata collection involves developing comprehensive plans and documenting protocols to ensure high-quality metadata[78]. This includes defining metadata variables, implementing standardized data collection procedures, enhancing sample diversity, and documenting all relevant details[36,38]. In addition, providing sufficient technical training can mitigate the expertise barrier, such as educating the use of software tools that track metadata on behalf of users, stamping workflows with software versions and provenance of annotations automatically[79–82].

The "Metadata for Machines" (M4M) workshops, part of the Three-point FAIRification Framework by GO FAIR and Research Data Alliance (RDA) members, represent a crucial initiative aimed at revolutionizing metadata practices in data-related communities[83,84]. The M4M workshops bring together domain experts and FAIR metadata specialists to collaboratively define and promote machine-actionable FAIR metadata components and templates. This effort is crucial for advancing the adoption of modular and extensible metadata schemas, promoting data interoperability and reuse. Additionally, the presence of data stewards within institutes could ensure researchers receive adequate training and facilitate effective data management practices[85]. For example, ETH Zurich Library launched the Data Stewardship Network (DSN) to foster collaboration among ETH employees engaged in research data management[86]. This initiative seeks to promote communication among data stewards regarding technical matters, enhance expertise in open research data (ORD) through training for both data stewards and ETH researchers, ensure adherence to ETH guidelines governing open research data practices, and provide educational materials and tutorials for effective research data management. In microbiome research, the NMDC focused on assessing and improving the adoption of community-driven metadata standards within the microbiome research community, aiming to understand and address barriers to adoption across diverse research domains, institutions, and funding agencies[70]. Workshops on ethical and legal aspects can educate researchers about their responsibility to use data for legitimate purposes, while also respecting and protecting individual privacy and confidentiality[87]. By investing in educational efforts, the scientific community can raise researchers' awareness of metadata sharing, foster a culture of standardized reporting, and improve data availability, accessibility, and quality[77,88].

**Funding agencies and journals: The pivotal roles of scientific journals and funding agencies in advancing and enforcing metadata sharing standards**

Funding agencies and journals play a crucial role in upholding and promoting guidelines. Journals contribute by mandating metadata and data sharing, establishing a standard reporting framework through requirements for authors to adhere to guidelines when submitting papers. For instance, journals like *Scientific Data, Nature* and *BMC Microbiome* have set examples by requiring researchers to disclose comprehensive metadata and data alongside their manuscripts. Despite proactive efforts by these journals to enhance metadata sharing practices, inconsistencies in compliance and enforcement persist. Addressing this challenge is essential for advancing the FAIR principles and fostering a more consistent and robust FAIR data ecosystem. Journals can ensure metadata consistency, completeness, and overall quality by mandating author submissions to adhere to established guidelines[21,71]. Journals can collaborate with data repositories, standard-setting organizations, or form consortiums with other publication groups to implement metadata standards[83]. The goal is to ensure every study in a participating journal incorporates the essential, standardized metadata fields. This enhances data searchability, reusability, and interoperability, ensuring consistent metadata structure for datasets published across multiple journals. This coordinated approach would facilitate data comparison, merging, and analysis across multidisciplinary or multi-journal studies, enhancing research transparency, replicability, and robustness.

Meanwhile, funding agencies can promote metadata sharing by requiring it as a condition for funding and incentivizing researchers to adopt and adhere to metadata reporting guidelines. A major funding agency like the National Institutes of Health (NIH) can play a pivotal role in establishing and promoting the widespread adoption of metadata reporting guidelines, which will help to create a more consistent and robust FAIR data ecosystem[89]. While the NIH has recently highlighted data management planning as a prerequisite for grant proposals, the absence of standardized metadata protocols hinders its mandatory inclusion. Additionally, recent NIH requirements for most research awards to include Data Management and Sharing plans may incentivize researchers to plan metadata sharing before generating data[90].

**Incentives and rewards: Driving forces for metadata availability**

The age-old "carrot vs stick" debate extends to the realm of metadata sharing. One approach involves providing researchers with the incentives and support they need to submit high-quality metadata, fostering a culture of voluntary compliance. Conversely, imposing penalties for non-compliance with metadata and data sharing guidelines risks discouraging researchers from submitting any data at all, potentially hindering scientific progress and limiting the availability of valuable research data. It is nevertheless essential to promote incentives that recognize the value of metadata sharing, such as acknowledging its contribution to research transparency, reproducibility, and data reuse[25]. For example, the publication of research papers in prestigious journals with high impact factor, which researchers actively strive to achieve, has the potential to incentivize and drive improvements in metadata reporting practices. This impact can be further amplified if these journals establish and enforce metadata reporting standards. In addition, the proliferation of data journals, platforms that mandate the use of standards-based metadata for omics datasets, presents a powerful opportunity to solidify metadata sharing standards[6,91]. Another potential solution to address the reluctance of researchers in sharing metadata, stemming from limited incentives, is to actively involve individuals who are already generating substantial amounts of data, particularly those who are comfortable sharing omics data. By engaging with data generators, we can collectively explore their insights and concerns, fostering collaborative brainstorming to develop effective strategies for enhancing metadata sharing. This collaborative approach aims to generate concrete ideas and actionable steps that will create a more conducive environment for comprehensive metadata sharing within the research community[92]. Furthermore, it is important to encourage other approaches, such as summary statistic level sharing, which can provide an alternative means of data sharing while still contributing valuable insights to the scientific data. By incentivizing metadata curation and mandating its reporting, we can harness the power of existing raw data to drive discovery and advance scientific knowledge.

**Improving infrastructures: Establishing a globally connected scientific community for metadata sharing with improved data security**

Establishing a robust infrastructure for sharing and storing metadata is essential for overcoming existing barriers and ensuring seamless integration with primary data[63]. Efforts should be directed towards promoting the development of secure data repositories that can accommodate large datasets while safeguarding data privacy. Robust data security measures and protocols must be implemented to mitigate the risks of data breaches and ensure the confidentiality of such metadata[93]. Implementing robust privacy safeguards, complying with legal requirements, and adhering to ethical guidelines will help mitigate risks and foster a trustworthy and ethically sound environment for the sharing of biomedical metadata[49,53,74,94]. The combination of physical separation and deliberate permalinking between metadata and data could improve data security and privacy. The strategy involves maintaining metadata devoid of re-identification elements, thereby reinforcing confidentiality. Simultaneously, the data, restricted in access to authorized individuals or algorithms, could potentially include sensitive details. By also actively accommodating these cultural considerations, it becomes possible to foster an environment that respects diverse perspectives while still advancing the broader goals of data sharing.

In addition, anonymization methods and federated analyses are two approaches that can be used to address data privacy concerns. Anonymization methods involve removing or obscuring personal information from data so that individuals cannot be identified. Federated analyses allow researchers to analyze data that is stored on different servers without having to share the data itself[95]. Both of these approaches can help to protect the privacy of individuals while still allowing researchers to conduct important research.

Additionally, researchers from research and academic institutions should be made aware of the value of metadata, and such institutes should allocate sufficient resources to support metadata management[51], including dedicating personnel and infrastructure to facilitate the annotation, documentation, and storage of metadata. Such institutions can also include line items in funding for these activities and positions. Adequate staffing levels and appropriate tools and technologies can streamline the metadata sharing process, minimizing delays and incompleteness. In order to effectively address gaps in metadata infrastructures across different countries, it is imperative to establish robust international collaborations and implement standardized protocols[96]. By sharing expertise and leveraging the strengths of each participating nation, a collaborative approach can help in developing comprehensive and efficient metadata storage solutions that transcend geographical boundaries.

**Discussion**

Despite the challenges, there are important opportunities to enhance the metadata availability. One key aspect is the provision of comprehensive training to personnel involved in data management, enabling them to effectively share metadata. This training would facilitate proficient metadata sharing practices, ensuring that valuable information is easily accessible and understandable. Web tools and other software solutions featuring user-friendly graphical user interfaces (GUI) can be developed to facilitate adherence to established metadata guidelines and alleviate the burden on researchers with limited computational skills. Furthermore, journals and public repositories can play the pivotal role to establish robust policies and guidelines that promote the dissemination of meticulous metadata, and foster transparency and standardization of data sharing practices. Funding agencies also play critical roles in incentivizing researchers to share standardized metadata, and further promote metadata availability. Furthermore, it is essential for the scientific community to develop and widely implement standards for metadata sharing. For example, the Minimum Information about a high-throughput SEQuencing Experiment[97] (MINSEQE) outlines the essential information required for the clear interpretation and reproducibility of high-throughput sequencing results. Similar to the MIAME guidelines for microarray experiments[21], adherence to MINSEQE enhances the integration of experiments across various modalities, maximizing the value of high-throughput research. This includes detailed information about the biological system, samples, and experimental variables, sequence read data, processed summary data, experiment and sample-data relationships, as well as essential experimental and data processing protocols. The collaborative effort between the scientific communities would ensure consistency and efficacy in data management practices, making it easier to locate and utilize relevant information across different research disciplines.

Improving the availability and quality of metadata brings numerous benefits to the scientific community and beyond[16,29,30,41,98,99], supporting data-driven decision-making and policy development across many fields[100], including healthcare, environmental sciences, and social sciences[16]. By providing comprehensive information, metadata empowers researchers, stakeholders, regulatory authorities, and the public to make informed choices based on reliable and relevant data. The study sheds light on the significant barriers that impede the sharing of metadata in scientific research. Acknowledging these formidable challenges takes on paramount importance, and doing so not only illuminates the current limitations but also lays the groundwork for prospective improvements. This proactive approach is essential for fostering a more conducive environment that facilitates the broader availability of metadata in the future, contributing to the advancement and transparency of scientific knowledge dissemination. Overall, investing in the improvement of metadata practices has wide-ranging benefits, fostering scientific progress, collaboration, reproducibility, and data-driven decision-making.

**Acknowledgements**
We thank Dr Walls for her valuable feedback and discussion. SM and Y-NH is supported by the National Science Foundation (NSF) grants 2041984, 2135954 and 2316223 and National Institutes of Health (NIH) grant R01AI173172. JHM is supported by grant U01-AG066833. MIL is supported by grant R01-HG009937. WAH is supported by (NIH) grants R01AG062547, R01AG082093, U19AG073172. AJB is supported by National Institute of Allergy and Infectious Diseases ImmPort Contract HHSN316201200036W, the UCSF Bakar Computational Health Sciences Institute, and the National Center for Advancing Translational Sciences of the National Institutes of Health (NIH) UL1 TR001872. LMS is supported by NIH/NHGRI 1U24HG012557-01.